\documentclass[twocolumn, superscriptaddress, amsmath, amssymb, aps]{revtex4-2}

\usepackage{graphicx}
\usepackage{dcolumn}
\usepackage{bm}
\usepackage{xcolor}
\usepackage{mathtools}
\usepackage{physics}
\usepackage[normalem]{ulem}
\usepackage{soul}
\usepackage{braket}
\usepackage{appendix}
\usepackage{xcolor}   
\usepackage{xr}
\usepackage{array} 

\usepackage{hyperref} 
\hypersetup{colorlinks=true, citecolor=blue}

\newcolumntype{P}[1]{>{\centering\arraybackslash}p{#1}}

\begin{document}
\title{Universal Statistics of Random Waves in Multi-Port non-Hermitian Scattering Systems}

\author{Nadav Shaibe}
\email{Corresponding author: nshaibe@umd.edu}
 \affiliation{Maryland Quantum Materials Center, Department of Physics, University of Maryland, College Park, Maryland 20742-4111, USA}

\author{Jared Erb}
 \affiliation{Maryland Quantum Materials Center, Department of Physics, University of Maryland, College Park, Maryland 20742-4111, USA}
 
\author{Steven M. Anlage}
 \affiliation{Maryland Quantum Materials Center, Department of Physics, University of Maryland, College Park, Maryland 20742-4111, USA}

\date{\today}

\begin{abstract}

We study the statistical properties of wave transport of a fixed multi-port non-Hermitian scattering system through wavefront shaping to explore all possible coherent illumination conditions.  This is a departure from traditional approaches utilizing cavity shaping to create an ensemble of systems with different detailed scattering arrangements that lead to uncorrelated field configurations. Using microwave measurements in the low channel-number limit, we empirically demonstrate universal statistics in the probability density functions (PDF) of reflectivity $\bm{r}$, transmissivity $\bm{t}$, and absorptivity $\bm{\alpha}$, Fisher information and the difference between forward propagating and time reversed transmissivity $\bm{t}-\bm{t}'$. Because the PDFs depend only on the number of input channels and the eigenvalues of Hermitian operators, we demonstrate how cavity shaping can be used to manipulate these eigenvalues, even creating Diabolic Point degeneracies, giving us direct control over the shapes and bounds of transport phenomena PDFs. We further investigate the PDFs when the excitations are finite bandwidth pulses as opposed to monochromatic waves, and determine the behavior in the limits of two channels and many channels. Our results establish a new paradigm in statistical investigations and cavity shaping applications by focusing on outcomes arising from all possible excitations, rather than creating conditions for one specific wavefront that optimally achieves a desired purpose.

\end{abstract}

\maketitle



\textit{Introduction}---The study of wave propagation in complex multi-scattering media and ray-chaotic billiards has been quantified through statistical descriptions of observable properties such as transmission, reflection, and conductance, informed by theoretical approaches based on random matrix theory (RMT) \cite{Dyson62,Brody81,BGS84,Been93,Li2017} and maximum entropy principles \cite{Mello85}. 
Historically these studies have been focused on creating a random ensemble of such systems that differ in detail but share the properties of a single universality class.  The extreme sensitivity of the wave properties to sub-wavelength boundary or scatterer perturbations serves to create a sampling of random wave configurations within their interior. 
Examples include reverberant chambers for electromagnetic compatibility studies with random plane waves \cite{Hill09,Serra17}, the use of tuning electrodes to change the shape of quantum dots to sample different values of conductance \cite{Marcus92}, the use of variable phase shifters to alter wave interference at the nodes of metric graphs \cite{Dietz2017,Rehemanjiang2018,Che2021,Che2022,Lawniczak2023,Lawniczak2024}, and the scrambling of scatterers in a wave scattering medium \cite{Zhang2010,Genack11}.  The statistical properties studied in such systems include eigenvalue \cite{Stock90,Graf92,Fyodorov2023,Fyodorov2024} and eigenfunction \cite{Chung2000,Kuhl2013,Ma2022_eig} statistical properties, as well as the properties of impedance and scattering matrices and derived quantities such as conductance \cite{Mendez2003,Fyodorov2004,Kuhl2005,Hemmady2005,Hemmady2005b,Hemmady2006,Fyodorov2005,Savin2006,Yeh2010,Yeh2010b,Lawniczak2010,Kumar2013,Nock2014,Lawniczak2019,Fyodorov2026}.  

In addition to variations in the structure of scattering systems, one can also employ variations in the frequency of injected waves, in the limit that the system is large compared to the wavelength and has complex ray trajectories \cite{Hill94,Henry06,Sameer05}.  This effective equivalence of shape-change and spectral variation is fundamental to many statistical studies of complex systems \cite{Ma2022}.  Other parametric quantities can be utilized to scramble wave properties as well, such as a magnetic field to alter the trajectories of charged particles within the system \cite{Beenakker1997,CastanedaRamirez2022,Zhang2023,erb2025}.

The time-honored paradigm is that single-realizations of a complex system will be dominated by system-specific and realization-specific details that are not of interest because they cannot be universal.  We now call this assumption into question and examine the statistical properties of single realizations of arbitrary scattering systems.  Remarkably, there are clear quantitative and universal statistical properties that have eluded experimental examination up to now, and we will present them here.

The statistical questions that we now ask concern the outcome of illuminating a given scattering system with all possible randomly-selected input waves, and considering the range of possible outputs in terms of reflection, transmission, and absorption.  In the past, wavefront shaping has been used almost exclusively for identifying one specific excitation that achieves a particular purpose or goal, such as maximizing transmission through a strongly-scattering system \cite{Vellekoop2008,Kim2012,Yu2013,Gerardin2014,Popoff2014,Cao22,Kostya23}, focusing or depositing energy at a location within a system \cite{Vellekoop2007,Cheng2014,Horstmeyer2015,Vellekoop2015,Durand2019,Cao24}, or maximizing absorption \cite{ChongCPA10,Wan2011,Baranov2017,Lei2020,Hougne2021_CPA,erb2027}. 

We test the prediction of Wang and Guo \cite{Wang2026}, which upsets this paradigm by using wavefront shaping to explore \textit{every} possible input excitation and examine the statistical properties of the corresponding output states. The prediction is that if an input wavefront $\bar{a}$ on $n$ channels results in an observable $o[\bar{a}]=\bar{a}^\dagger O\bar{a}$, $O$ being a Hermitian operator, then the probability density function of $o$ is a B-spline piecewise polynomial of degree $n-2$ with knots at the eigenvalues $\lambda$ of $O$. We find that the predicted distribution functions are universal, in the sense that their shape is independent of the microscopic configuration and global properties of the system, including the degree of loss (incorporating non-Hermitian scattering systems). Further, these predictions hold for systems both large compared to the wavelength (such as those studied through ensemble statistics), and small compared to the wavelength.

\textit{Results}---Our experiments are conducted in the microwave scattering regime, where we have the advantage of easily accessing the low channel-number limit, where the PDF of any observable most dramatically change with $n$ and are not well approximated by a Gaussian, as opposed to what is seen in the many channel limit \cite{Davy2013}. We explore the predicted PDFs using microwave graph networks with different topologies \cite{Kottos1997,Hul2004,Hul2005,Lawniczak2009,Lawniczak2012,Dietz2017,Lei2020,Che2021,Che2022,shaibe2025}, and a ray-chaotic quarter bowtie billiard \cite{So1995,Wu1998,Chung2000,Ma2022,Erb2024,erb2025,Erb2026,shaibe2027,erb2027}. The $L\times L$ scattering matrices we measure can be block-decomposed as $S=\begin{bmatrix}\rho & \tau' \\\tau & \rho'\end{bmatrix}$, where $\rho$ is $n\times n$ and corresponds to reflection at the ``input'' channels, and $\tau$ which is $m \times n$ and corresponds to transmission from the ``input'' channels to the ``output'' channels. In our experimental systems, all the channels are treated on equal footing, so which ones are used for input or output is an arbitrary choice. For a given system, we can vary $n$ from $2$ to $L$, or keep $n$ fixed and change which channels are in the set of inputs.


  \begin{figure*} [thb]
	\centering
	\includegraphics[width=.95\textwidth]{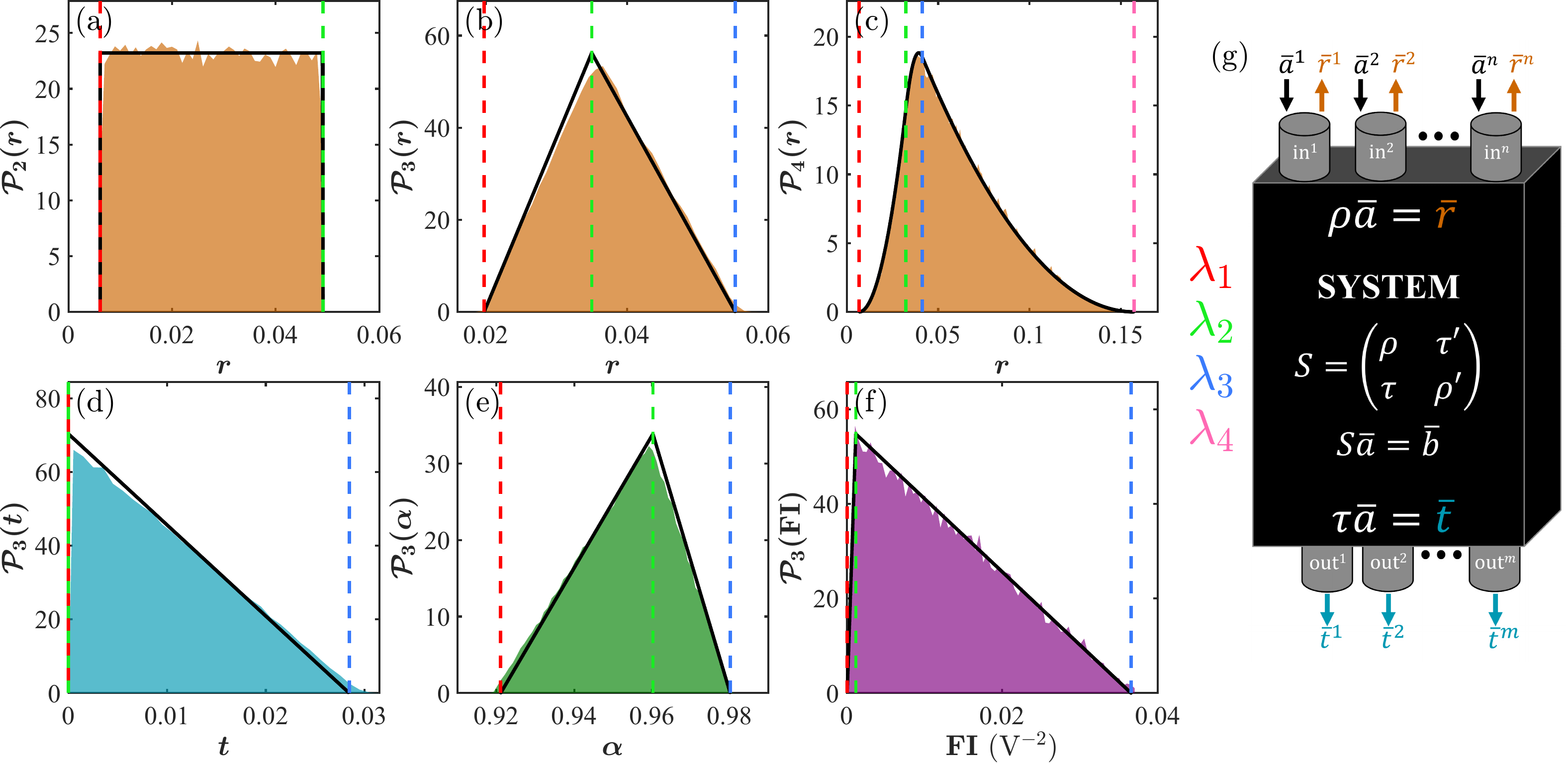}
	\caption{(a-c) Experimentally determined PDFs (orange) of reflectivity $\bm{r}$ from an $L=4$ port microwave billiard in the cases of number of input ports $n=[2,3,4]$. Black curves depict theoretical prediction of B-spline (piecewise polynomial) with knots at the eigenvalues of $R=\rho^\dagger\rho$. The location of the eigenvalues are marked in vertical dashed lines in ascending order: red, green, blue, pink. (d-e) PDFs of transmissivity $\bm{t}$ (blue) and absorptivity $\bm{\alpha}$ (green) from the same measurement as panel (b). (f) Experimental PDF of Fisher information $\mathrm{FI}$ (purple) of a voltage applied to a metasurface embedded within the $L=3$ port microwave billiard. For all panels, the billiard was measured at $8~\mathrm{GHz}$. (g) schematic of measurement process.}
	\label{FIG_1}
\end{figure*}

Measurements are performed using an $L\leq6$ channel, single source Vector Network Analyzer (VNA), and another $L\leq4$ channel $n\leq4$ source VNA. With the first VNA, we can measure scattering matrices $S$ that are up to $6\times6$. With the second VNA, we can measure $S$ up to $4\times4$, but we can also inject a wavefront $\bar{a}$ with independent amplitude and phase control on $n\leq4$ channels. We then measure the amplitude and phase of waves leaving the cavity on all $L=n+m$ channels, with the wavefront returning on the $n$ input channels being $\bar{r}=\rho\bar{a}$ and the wavefront transmitted to the $m$ ``output'' channels is $\bar{t}=\tau\bar{a}$. The scalar reflectivity $\bm{r}[\bar{a}]$, transmissivity $\bm{t}[\bar{a}]$, and absorptivity $\bm{\alpha}[\bar{a}]$ of a wavefront are calculated via:
\begin{align}
     \bm{r}[\bar{a}]=\frac{\bar{r}^\dagger\bar{r}}{\bar{a}^\dagger\bar{a}}=\frac{|\bar{r}|^2}{|\bar{a}|^2},\nonumber \\ \newline
    \bm{t}[\bar{a}]=\frac{\bar{t}^\dagger\bar{t}}{\bar{a}^\dagger\bar{a}}=\frac{|\bar{t}|^2}{|\bar{a}|^2},\nonumber\\ \newline
    \bm{\alpha}[\bar{a}]=1-r[\bar{a}]-t[\bar{a}]. \nonumber
\end{align}
A schematic of this measurement process is shown in Fig.~\ref{FIG_1}(g), with system details suppressed. See Supp. Mat. section~\ref{Sup_Sys} for a complete discussion and schematics of the systems measured in this work. To arrive at an estimate of the PDF $\mathcal{P}_n$ of these quantities, we repeat the injection process for approximately $100,000$ input wavefronts with amplitudes randomly chosen from the Dirichlet distribution and phases randomly chosen from a uniform distribution.

Figure \ref{FIG_1} shows statistical PDFs for resulting output quantities measured for a fixed configuration of the microwave billiard at a fixed frequency for a broad set of input wavefronts.   In Fig.~\ref{FIG_1}(a-c), we show that the experimentally determined PDFs of reflectivity $\bm{r}$ (orange areas) match the prediction of a B-spline (black curve) with knots at the eigenvalues of the reflectance matrix $R=\rho^\dagger\rho$ (colored dashed lines) for the cases of $n=[2,3,4]$ \cite{Wang2026}. For brevity, we only show the PDFs of transmissivity $\bm{t}$ and absorptivity $\bm{\alpha}$ for the case of $n=3$ in panels (d) and (e), which come from the same measurement as panel (b). Panel (d) shows that the transmittance matrix $T=\tau^\dagger\tau$ has two degenerate eigenvalues at $0$. This is due to there being $n$ eigenvalues but for $T$, at most only $m$ eigenvalues can be non-zero and in this case $m=1$. Notably, this Diabolic Point degeneracy \cite{Yang2023,Chen2025} skews the PDF, and in non-trivial cases ($m\geq n$) we can create such degeneracies to further control the PDFs, which we demonstrate in the Supp. Mat. Sec.~\ref{Sup_DP}. Panel (e) demonstrates that the billiard is a highly absorptive system, in which every possible incident wavefront is over $90\%$ absorbed. Measured $P_n(\bm{\alpha})$ PDFs from a much lower loss microwave graph are available in Supp. Mat. Sec.~\ref{Sup_EXP}.

In panel (f), we demonstrate the universality of this result beyond the simple transport observables of $\bm{r}$,$\bm{t}$, and $\bm{\alpha}$. We do this by tuning the voltage applied to a globally biased varactor diode-loaded metasurface \cite{Sleasman2023,Erb2024} between $V_0+\delta V$ and $V_0-\delta V$. We then form the Fisher information operator (FIO) $F_V = \left( \frac{\partial S}{\partial V} \right)^\dagger\frac{\partial S}{\partial V}$ \cite{Bouchet2021,shaibe2027}. The Fisher information returned by a wavefront $\bar{a}$ can be experimentally determined from
\begin{equation}
    \bm{\mathrm{FI}}[\bar{a}]=\left|\frac{\partial \bar{b}}{\partial V}\right|^2/|\bar{a}|^2, \nonumber
\end{equation}
where we calculate $\frac{\partial \bar{b}}{\partial V}$ from the finite difference $\frac{\bar{b}|_{V+\delta V}-\bar{b}|_{V-\delta V}}{2\delta V}$. If measurement of the full scattering matrix is not possible, the Fisher information operator can also be formed using $\rho$ or $\tau$, and the $\bm{\mathrm{FI}}$ calculated from $\bar{r}$ and $\bar{t}$.  The PDF of $\bm{\mathrm{FI}}$ shown in Fig.~\ref{FIG_1}(f) also agrees with the predicted universal B-spline form.

It is known that the eigenvector associated with the largest eigenvalue of the Fisher information operator $F_x$ is the optimal input for maximizing the returned information about the associated parameter $x$ \cite{Bouchet2021,shaibe2027}. In this billiard, we find the that there are orders of magnitude between the eigenvalues of $F_V$ ($\lambda_1=5.7\times10^{-5}~\mathrm{V}^{-2},\lambda_2=1.2\times10^{-3}~\mathrm{V}^{-2}$, and $\lambda_3=3.7\times10^{-2}~\mathrm{V}^{-2}$). If we ask how much information will be returned by a random input wavefront, we see that for this situation there is an $83.8\%$ probability of the returned $\mathrm{FI}$ being within an order of magnitude of the optimal amount. In Ref.~\cite{shaibe2027}, it was found that the optimal wavefront was not required to achieve certain tasks, just a wavefront associated with Fisher information above some threshold, which for this same billiard and metasurface was within an order of magnitude of the largest $F_V$ eigenvalue. These two results together suggest that an arbitrary wavefront could be used with good odds of accomplishing the desired task. However, this would not be true in the case of large $n$ as the bulk of the PDF becomes solely defined by the interior eigenvalues \cite{Wang2026}, and so most wavefronts would return $\mathrm{FI}$ much less than the largest $F_V$ eigenvalue, and the useful wavefronts would reside in the tail of the probability density function.

More observables that display PDFs with this universal behavior are provided in Appendix~\ref{App_Uni}, with a figure demonstrating some examples. We emphasize at this point that these experimental verifications of the theory in Ref.~\cite{Wang2026} demonstrate that knowledge of only the eigenvalues of a Hermitian operator is sufficient to completely describe the statistical outcome of wave excitations, but does not determine what happens to a particular wavefront.


Past work on cavity shaping has been mostly limited to creating conditions required for a specific wavefront to demonstrate various phenomena, such as reflectionless scattering \cite{Sweeney2020,Imani2020,Stone2020,Sol2023,Jiang2024}, Coherent Perfect Extinction \cite{Guo2023,Alhulaymi2025}, and Coherent Perfect Absorption \cite{Pichler2019,Frazier2020,Hougne2021_CPA,Erb2024,erb2025,Erb2026,erb2027}, etc. Some works have demonstrated CPA for all wavefronts with a single cavity, but this utilizes a designed, degenerate cavity that is not reconfigurable \cite{Slobodkin2022,Horner2024}. With perfect knowledge of the statistical outcome of all wave excitations deduced from knowledge of the scattering matrix $S$, we introduce a new paradigm of cavity shaping based on optimizing the PDF of the observable of interest such that a majority of input wavefront exhibits a desired behavior. We provide an example of this new paradigm in the context of minimizing and maximizing the absorption experienced by all waves entering a scattering system.

\begin{figure} [thb]
	\centering
	\includegraphics[width=\columnwidth]{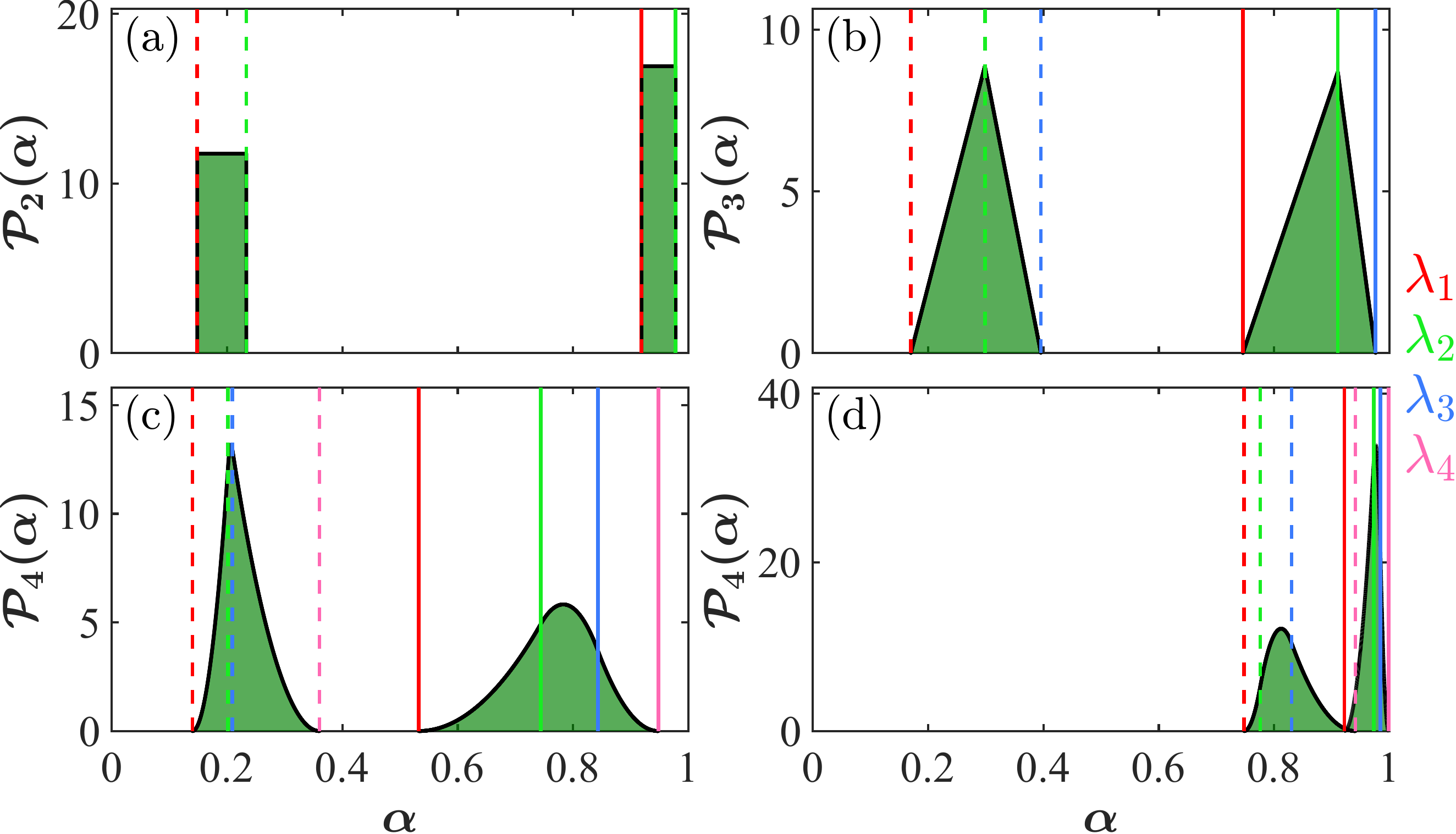}
	\caption{Absorptivity $\bm{\alpha}$ PDFs demonstrating the new cavity shaping paradigm. Panels (a-c) are from microwave Petersen graphs with $L=[2,3,4]$ channels. In each panel, both PDFs are formed from data taken at the same frequencies ($7.106~\mathrm{GHz}$,~$7.791~\mathrm{GHz}$, and $7.136~\mathrm{GHz}$ respectively), with the only differences being small local perturbations to the lengths of some of the bonds of the graph such that the overall length is fixed. One of the PDFs shows the lowest average absorptivity measured while the other shows the highest. (d) Data from the $L=4$ port microwave billiard at $8~\mathrm{GHz}$ where the average absorptivity was maximized and minimized through an optimization function.}
	\label{FIG_2}
\end{figure}

Panels (a-c) of Fig.~\ref{FIG_2} are PDFs of absorptivity $\bm{\alpha}$ that come from experimental ensembles of microwave networks with the topology of the Petersen graph. Each individual realization of the graph in these ensembles has the same total electrical length of $3.92~\mathrm{m}$, with the differences coming from small (millimeter to single centimeter) changes to six of the fifteen bond lengths accomplished through programmable analog phase shifters \cite{shaibe2025,Shaibe_2025_Top}. Therefore the uniform attenuation rate of each realization should be the same, with only the details of the wave interferences at the nodes of the graph changing. We see from the pairs of distributions in Fig.~\ref{FIG_2} that just by making these small changes, we are able to, at arbitrary frequency, completely change the degree to which all waves are absorbed.

Panel (a) of Fig.~\ref{FIG_2} comes from an $L=2$ port Petersen graph at $7.106~\mathrm{GHz}$. The two solid green areas show the B-spline PDF of $\bm{\alpha}$ for the realization with the smallest average $\bm{\alpha}$ and largest average $\bm{\alpha}$. Panels (b) and (c) show the same things but for an $L=3$ port Petersen graph at $7.791~\mathrm{GHz}$ and an $L=4$ port Petersen graph at $7.136~\mathrm{GHz}$, respectively. Note that an optimization algorithm was not utilized to arrive at these cavity configurations, they are simply random settings of the phase shifters taken from large ensembles used for traditional cavity-shaping statistical analysis. In contrast, panel (d) shows two PDFs of $\bm{\alpha}$ from the quarter bowtie billiard at $8~\mathrm{GHz}$ for which the average $\bm{\alpha}$ was optimized using a surrogate optimizer to have the smallest and largest value possible \cite{erb2027}. In the case of largest average $\bm{\alpha}$, we were able to configure the system such that there was a $90\%$ probability of a random wavefront being at least $95\%$ absorbed. However, because the billiard is so absorptive, the dynamic range of the absorptivity matrix ($A=1-\rho^\dagger\rho-\tau^\dagger\tau$) eigenvalues is much smaller with a hard lower bound, and even with optimization we were unable to get any configuration that had any waves that would be absorbed less than $70\%$.

In Supp. Mat. Sec.~\ref{Sup_DP}, we also show PDFs from the same ensembles and frequencies as panels (a-c) where instead of looking at the average $\bm{\alpha}$, we consider the variance of $\bm{\alpha}$. These results show the utility and adaptability of this new form of cavity shaping, as not just the average or variance, but also the bounds (smallest and largest eigenvalue) on any observable can be controlled. We also show examples of cavity shaping to create degeneracies in the eigenvalues which are Diabolic Points (DP), not Exceptional Points, because the operators are Hermitian. These are actually easier and more common to create than scattering singularities such as CPA \cite{Shaibe_2025_Top,Alhulaymi2025,erb2027}, which further argues that cavity shaping to control scattering outcome PDFs is advantaged over standard methods. There are many relevant tasks, such as maximizing or minimizing a transport observable \cite{Vellekoop2008,Kim2012,Yu2013,Gerardin2014,Popoff2014,Cao22,Kostya23,Sol2023,Jiang2024,Pichler2019,Frazier2020,Hougne2021_CPA,Erb2024,erb2025,Erb2026,erb2027} or focusing energy at a location \cite{Ambichl2017,Bouchet2021,Hougne2021_Store,Cao24,shaibe2027}, for which the optimal wavefront is the eigenvector of a Hermitian operator associated with it's largest eigenvalue. For these tasks, we suggest cavity shaping such that a high percentage, if not all incident wavefronts, can accomplish the task within an acceptable tolerance. Taken a step further, it should be possible to cavity shape a system such that all inputs achieve multiple desired objectives, such as specified values of reflectance, transmittance, and absorptance simultaneously, as was done for a single input in a fixed system in Ref.~\cite{Li2025}.

There are, however, important limitations on what can be accomplished with passive complex scattering systems \cite{Guo2024}, and one already discussed is the degree of homogeneous loss of the scattering system. Intrinsic homogeneous loss puts a lower bound on $\bm{\alpha}$ and an upper bound on $\bm{r}$ and $\bm{t}$ (we predict also on $\mathrm{FI}$), whereas inhomogeneous loss in the form of an attenautor, lumped element, or unmonitored open channel, can be avoided to varying degrees by different wavefronts. In general, lumped loss could lead to a wider dynamic range for the absorptivity matrix $A$ eigenvalues. The coupling strength of the channels to the cavity modes can also limit the effectiveness of any cavity shaping efforts, as imperfect coupling puts a lower bound on $\bm{r}$ (and consequently an upper bound on $\bm{t}$ and $\bm{\alpha}$).


Now we consider the case of exciting a scattering system with a finite bandwidth pulse as opposed to a monochromatic wave. The resulting PDF of any observable becomes the sum of the PDFs at each frequency within the bandwidth. We highlight the extreme case of $n=2$ and then discuss what happens as $n$ increases.

\begin{figure} [thb]
	\centering
	\includegraphics[width=\columnwidth]{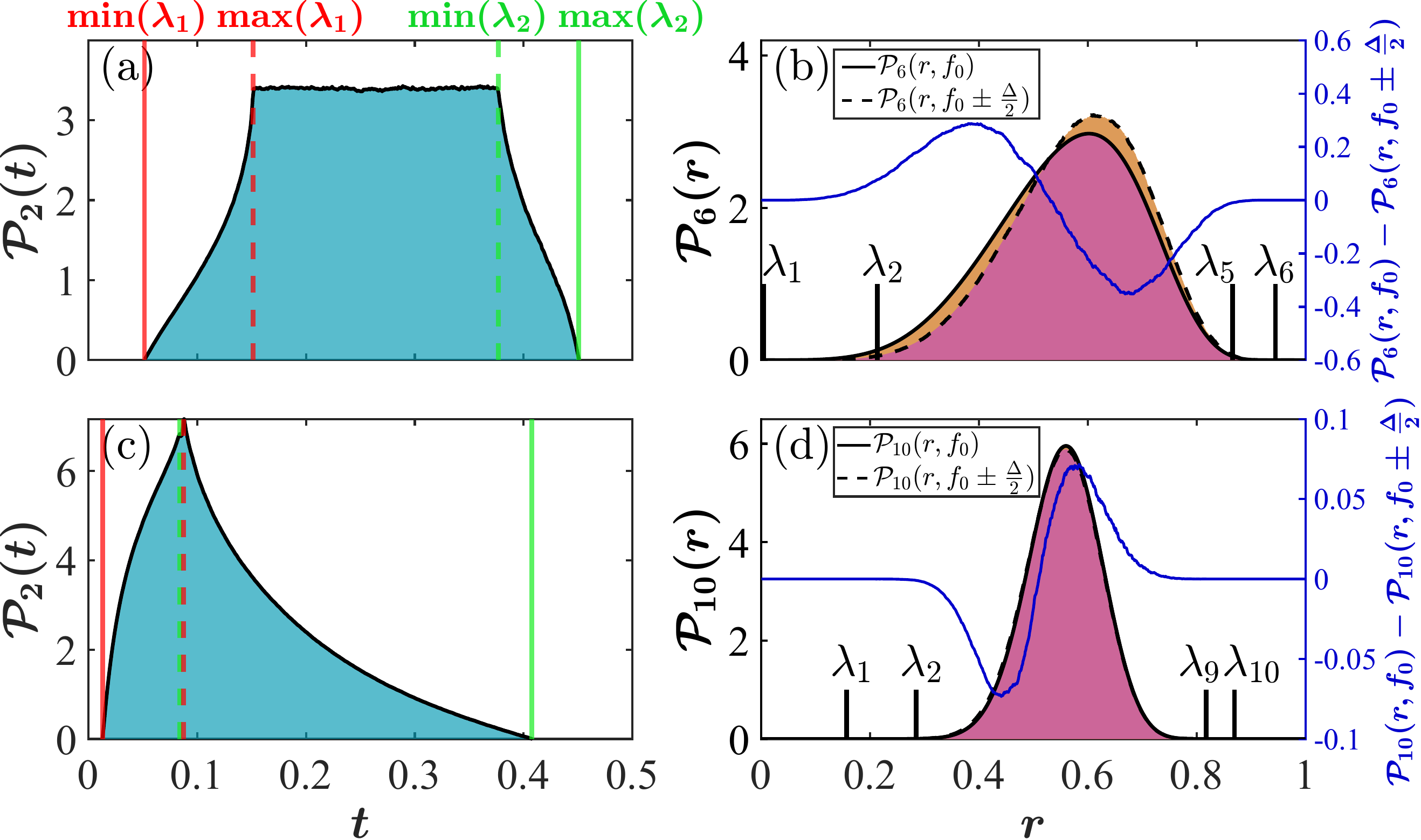}
	\caption{(a) and (c), PDFs of transmissivity $\bm{t}$ for two input channels $n=2$  and $L=6$ channel $K_{3,3}$ complete bipartite microwave graph where the input waves $\bar{a}$ include frequencies covering a complete mean mode spacing $\Delta$ centered around $f_0=15.55~\mathrm{GHz}$. (b) six input channel $n=6$ PDF of reflectivity $\bm{r}$ for the same graph as panel (c), showing the change from single frequency $f_0$ (solid curve) to a bandwidth of $\Delta$ (dashed curve). The overlapping region of the two PDFs is colored pink, and the difference is plotted as a blue curve. (d) ten input channel $n=10$ PDF of reflectivity $\bm{r}$ from an RMT simulation, demonstrating that as $n$ gets larger, the change from single frequency to bandwidth PDFs becomes negligible.}
	\label{FIG_3}
\end{figure}

For $n=2$, the PDF of any observable produced by a monochromatic wave has been shown to be a uniform distribution. For a finite bandwidth pulse, we find that there is still a region of the PDF that is uniform, bounded below (above) by the largest (smallest) value of $\lambda_1$ ($\lambda_2$) in the frequency band. If $\max[\lambda_1]\geq\min[\lambda_2]$, then the PDF develops a sharp point, and no uniform region remains. These two cases are shown in Fig.~\ref{FIG_3}, panels (a) and (c), both of which have PDFs of $\bm{t}$ from an experimental $L=6$ channel, six node complete bipartite graph $K_{3,3}$, with the bandwidth of the pulse being one mean mode spacing $\Delta=38.46~\mathrm{MHz}$. The center frequency for both is $15.55~\mathrm{GHz}$, but the graph itself has been modified between the two panels through the use of programmable analog phase shifters to change multiple bond lengths.

As the number of input channels $n$ grows, the effect of having a finite bandwidth excitation lessens. In panel (b), we show the single frequency PDF of $\bm{r}$, $\mathcal{P}_6(\bm{r},f_0)$, outlined in a solid curve from the same graph and center frequency as panel (c), and the $\Delta$ frequency bandwidth PDF $\mathcal{P}_6(\bm{r},f_0\pm\frac{\Delta}{2})$ outlined in a dashed curve. Both PDFs are colored orange, and their overlap is represented in pink. The difference $\mathcal{P}_6(\bm{r},f_0)-\mathcal{P}_6(\bm{r},f_0\pm\frac{\Delta}{2})$ is shown as a blue curve corresponding to the right y-axis. Panel (d) shows a similar plot, but this time taken from an $L=10$ channel Random Matrix Theory (RMT) generated scattering matrix, using the algorithm outlined in Ref.~\cite{Shaibe_2025_Top}. Already at $n=10$, the single frequency and finite bandwidth PDFs are nearly indistinguishable. Since the single frequency PDFs converge to a Gaussian in the large $n$ limit, and the sum of Gaussians is also a Gaussian, we predict that finite bandwidth PDFs at large $n$ converge to the PDF of the center frequency of the excitation band.


\textit{Conclusions}---In this Letter, we have emprically demonstrated universal statistical features in the observables of a fixed system, establishing a new paradigm in the statistical study of multi-port Non-Hermitian scattering systems. Our work is in the interesting low-$n$ limit, and we demonstrate universality in transmissivity $\bm{t}$, reflectivity $\bm{r}$, absorptivity $\bm{\alpha}$, as well as generalizing to other observables such as Fisher information. We not only validate the predictions of Wang and Guo \cite{Wang2026}, but further generalize them to finite bandwidth excitations, and show the effects on the PDFs in the limits of $n=2$ and large $n$. We show how controlling the eigenvalues of a Hermitian operator through cavity shaping methods can completely change how the system responds to the set of \textit{all} input wavefronts. With clever tuning, including controlling the extremal eigenvalues, manipulating the eigenvalue mean, and creating Diabolic Point degeneracies, one can create a situation in which most, if not all, incident wavefronts yield the desired response. These results provide guidance for new cavity and medium shaping techniques and objective functions.



\bigskip
\textbf{Acknowledgements} 
We are grateful to Prof. Cheng Guo (UT Austin) for valuable discussions. 
This work was partially supported by NSF/RINGS under grant No. ECCS-2148318, ONR under grant N000142312507, and DARPA WARDEN under grant HR00112120021.   



\bigskip


\appendix

\section{Universality Extended to Other Observables} \label{App_Uni}

In Ref.~\cite{Wang2026}, Wang and Guo apply their theory to $\bm{t}$, $\bm{r}$, and $\bm{\alpha}$, but claim it extends naturally to other measured observables. We demonstrated that this is true for Fisher information, and in this section suggest more observables with statistics described by the B-spline prediction. First, we consider how non-reciprocal or time-reversed transmission differs from  the ``forward propagating'' transmission in the case of $n=m$.  This is related to the concept of transmission time delay difference introduced in Refs.~\cite{Lei2024,shaibe2025}, but generalized to sub-blocks of the scattering matrix as opposed to individual elements. By applying the identical wavefront $\bar{a}$ on the $n$ input channels and then the $m$ output channels, we can measure the ``forward'' transmissivity $\bm{t}$ and the ``reversed'' transmissivity $\bm{t}'$ and calculate the difference $\bm{t}-\bm{t}'$.

\begin{figure} [thb]
	\centering
	\includegraphics[width=\columnwidth]{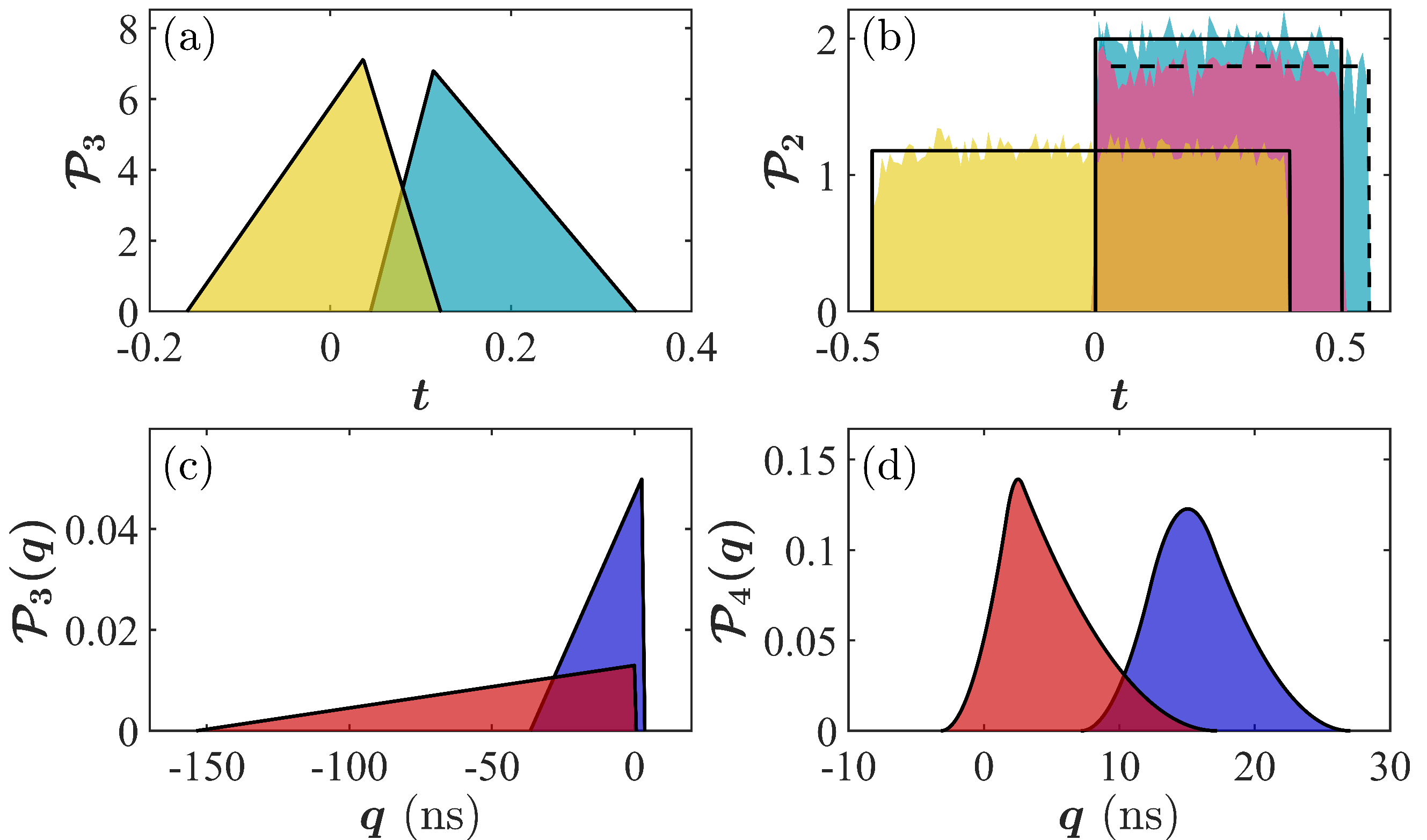}
	\caption{(a) PDF of transmissivity $\bm{t}$ (teal) and of non-reciprocal / difference from time reversed transmissivity $\bm{t}-\bm{t}'$ (yellow) from the same graph and at the same single frequency $f=15.55~\mathrm{GHz}$ as Fig.~\ref{FIG_3}(a-b). (b) Experimental demonstration of non-reciprocal transmissivity in a $4$ channel Petersen graph incorporating a microwave circulator that is $2.38$ meter long measured at $1.2715~\mathrm{GHz}$. PDFs of $\bm{t}$ and $\bm{t}'$ are indicated by the light blue areas bounded by the solid and dashed black curves, their overlap is marked in pink, and $\bm{t}-\bm{t}'$ is marked in yellow. (c-d) PDFs of real (blue) and imaginary (red) time delays $\bm{q}$ experienced by a wavefront $\bar{a}$ injected into the $3$ and $4$ channel Petersen graphs from Fig.~\ref{FIG_2}. }
	\label{FIG_App1}
\end{figure}

Fig.~\ref{FIG_App1}(a) shows $\mathcal{P}_3(\bm{t})$ from the same graph as Fig.~\ref{FIG_3}(c) in blue ($\mathcal{P}_3(\bm{t}')$ is identical), and $\mathcal{P}_3(\bm{t}-\bm{t}')$ in yellow. This graph was perfectly reciprocal ($S_{xy}=S_{yx}$), yet for a generic wavefront $\bar{a}$ we see a range for the difference $\bm{t}-\bm{t}'$. Only in the trivial case of $n=m=1$ would $\bm{t}$ and $\bm{t}'$ be generically identical. However, we do find that for reciprocal systems the mean of $\mathcal{P}_n(\bm{t}-\bm{t}')$ will always be $0$, though as shown, it does not necessarily have to be symmetric about zero. Panel (b) shows experimental data where we performed the measurement on a $2.38$ meter long non-reciprocal microwave network with the topology of the Petersen graph with $n=m=2$. The blue areas are the PDFs of $\bm{t}$ and $\bm{t}'$, their overlap shown in pink, and the yellow area is $\mathcal{P}_2(\bm{t}-\bm{t}')$. Because this graph is non-reciprocal, as can be inferred from $\mathcal{P}_2(\bm{t})\neq\mathcal{P}_2(\bm{t}')$, the mean of $\mathcal{P}_2(\bm{t}-\bm{t}')$ is not zero, but instead $-0.02$. One can also consider $\bm{r}-\bm{r}'$, though in general no relationship between $\rho$ and $\rho'$ can be assumed so no special features, such as a mean of zero, can be attributed to $\mathcal{P}_n(\bm{r}-\bm{r}')$.

In Hermitian scattering, the Wigner-Smith operator (WSO): $Q=-iS^{-1}\frac{\partial S}{\partial\omega}$ \cite{Wigner1955,Smith1960,Brouwer1997,Cunden2016}, and the generalized Wigner-Smith operator (GWSO): $Q_\theta=-iS^{-1}\frac{\partial S}{\partial\theta}$ \cite{Ambichl2017,Horodynski2020,Hougne2021} are Hermitian operators, and the their respective observables of time delay $\bm{q}$ \cite{Smilansky2017,Smilansky2018} and momentum $\bm{q}_\theta$ (when $\theta$ is a displacement) \cite{Orazbayev2024,Nerson2026} can be measured for an ensemble of wavefronts $\bar{a}$. B-Spline PDFs for $\bm{q}_\theta[\bar{a}]$ were suggested in the Supplemental Material of Ref.~\cite{Nerson2026}, and it is not unreasonable to expect the same for $\bm{q}[\bar{a}]$.

Perhaps more interestingly, in non-Hermitian scattering where $Q$ and $Q_\theta$ are not Hermitian, the real part of time delay can become negative, and there is an imaginary component as well \cite{Asano2016,Lei2021,Chen_Ring22,shaibe2025}. We then predict that the PDFs of $\mathrm{Re}[\bm{q}]$ and $\mathrm{Im}[\bm{q}]$ will each individually have the B-spline form, and these are quantities that can be independently measured in the time domain through pulse propagation \cite{Giovannelli2025,Giovannelli2026}. In panels (c) and (d) of Fig.~\ref{FIG_App1}, we show the expected PDFs of $\mathrm{Re}[\bm{q}]$ and $\mathrm{Im}[\bm{q}]$ for the $L=3$ and $L=4$ channel Petersen graphs featured in Fig.~\ref{FIG_2}. Note that the values of $\bm{q}$ are truly unbounded, though we know there is a $-3$ power-law suppression of extreme positive and negative values of both $\mathrm{Re}[\bm{q}]$ and $\mathrm{Im}[\bm{q}]$ \cite{shaibe2025}. While the interpretation of the complex GWSO and what the observable can be measured is less well understood, we expect the PDFs of $\mathrm{Re}[\bm{q}_\theta]$ and $\mathrm{Im}[\bm{q}_\theta]$ to also share the B-spline form.


\newpage

\newpage

\setcounter{figure}{0}
\setcounter{equation}{0}
\setcounter{section}{0}
\makeatletter
\renewcommand{\figurename}{Fig.}
\renewcommand{\thefigure}{S\arabic{figure}}   

\renewcommand{\tablename}{Table}
\renewcommand{\thetable}{S\Roman{table}}    

\renewcommand{\theequation}{S\arabic{equation}}    

\renewcommand{\thesection}{S\arabic{section}}
\makeatother

\begin{center}
\textbf{UNIVERSAL STATISTICS OF RANDOM WAVES IN MULTI-PORT NON-HERMITIAN SCATTERING SYSTEMS}
\vspace{0.4cm}
\\
SUPPLEMENTARY MATERIAL 
\vspace{0.4cm}



\vspace{0.4cm}
Nadav Shaibe, Jared M. Erb, and Steven M. Anlage

\end{center}

\section{Experimental Details} \label{Sup_Sys}

\begin{figure} [thb]
	\centering
	\includegraphics[width=\columnwidth]{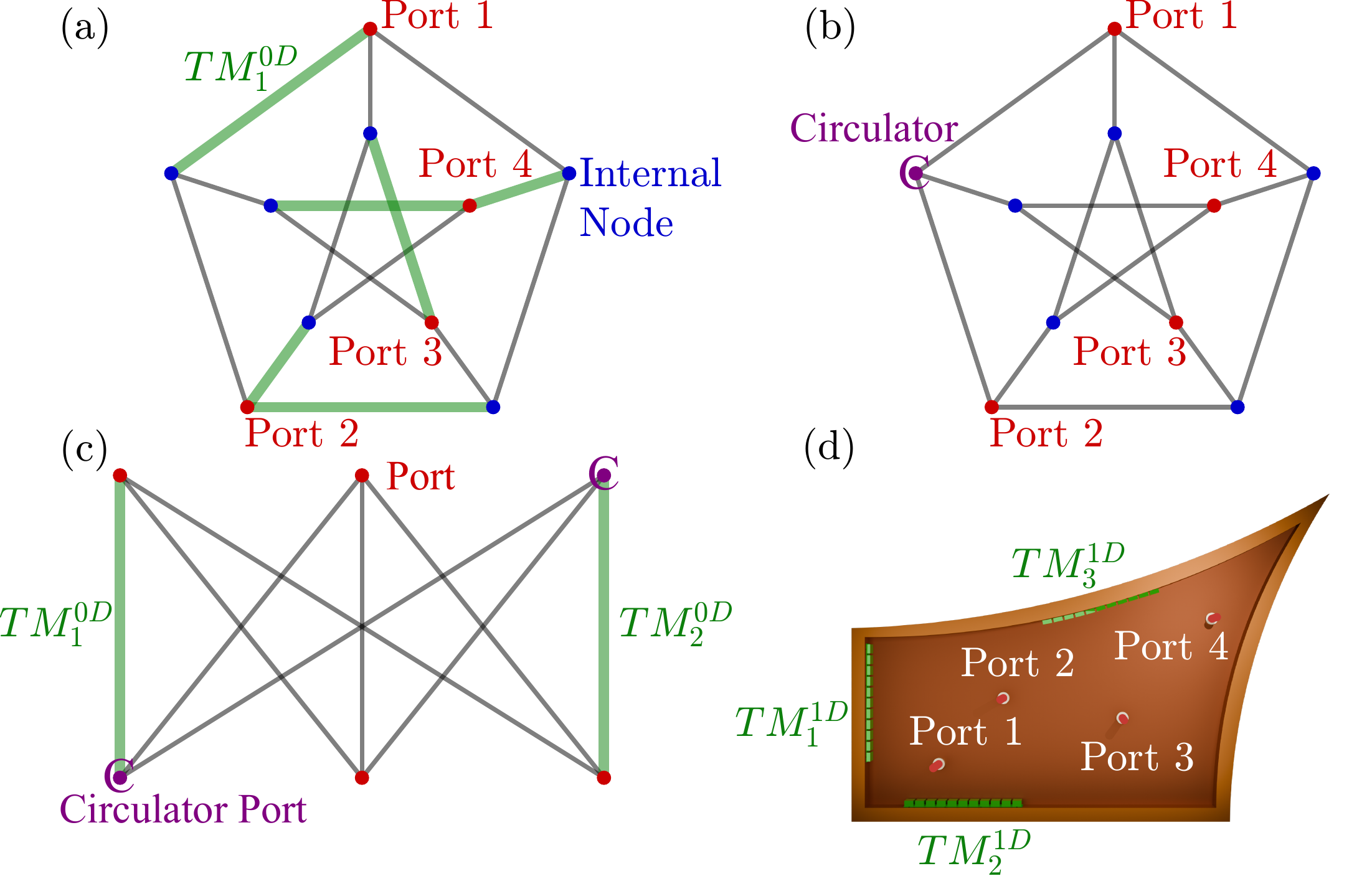}
	\caption{(a) Schematic of Petersen graph with 6 variable phase shifters (green lines), one of which is labeled $TM_1^{0D}$. The internal nodes are labeled as blue dots, and the port nodes connected to the VNA are labeled with red dots. (b) Schematic of Petersen graph with a circulator on one of the nodes, marked in purple, which introduces scattering non-reciprocity. (c) Schematic of the $K_{3,3}$ bipartite graph with two varibale phase shifters labeled in green and two of the ports having circulators. (d) Schematic of quarter bowtie billiard with three metasurfaces labeled $TM_1^{1D}$, $TM_2^{1D}$, and $TM_3^{1D}$.}
	\label{FIG_Sup_Sys}
\end{figure}

In this section, we provide a more complete discussion of the multiport wave systems measured, schematics for which are provided in Fig.~\ref{FIG_Sup_Sys}. Various different microwave graphs were used in this work, but all were built using standard coaxial SMA cables for the bonds and three way tee-junctions for the nodes. To connect a node of the graph to the Vector Network Analyzer (VNA), we attach two tee-junctions together to make a four way junction. Two different graph topologies were used, the first being the Petersen graph and the second being the $K_{3,3}$ bipartite graph.

The Petersen graph has ten vertices and fifteen edges. It is useful for our purposes as it is a cubic graph (each vertex having degree three) which we can build with tee-junctions, and it is the smallest cubic graph that can be formed such that we can separate the nodes into ``internal nodes'', ones not connected to the VNA, and ``port nodes'', connected to the VNA, and not have any port node directly connect to any other port node. This gives more opportunities for wave interference before the excitations leave the system, providing greater tunability and more complex scattering. A schematic with six variable phase shifters as was used for Figs.~\ref{FIG_2} and \ref{FIG_App1}(c-d) is shown in Fig.~\ref{FIG_Sup_Sys}(a), and panel (b) has a schematic of the Petersen graph with a circulator on one of the nodes to break reciprocity, which was used for Fig.~\ref{FIG_App1}(b). Note that while we label four ports in panels (a) and (b), some of the Petersen graph measurements were done with two or three ports. For those measurements, the extra unused port nodes are replaced with single tee-junctions are become internal nodes.

The $K_{3,3}$ graph is the complete bipartite graph with six vertices and nine edges. The vertices can be separated into two sets, $P$ and $Q$, such that a vertex in set $P$ is connected to all the vertices in set $Q$ and to no other vertices in $P$. The data in Figs.~\ref{FIG_3}(a-c) and \ref{FIG_App1}(a) comes from a $K_{3,3}$ graph with two phase shifters ($TM_1^{0D}$ and $TM_2^{0D}$) and $L=6$ ports. Two of the ports have circulators instead of tee-junctions. A schematic of this graph is shown in Fig.~\ref{FIG_Sup_Sys}(c).

We also measure a ray-chaotic quarter bowtie billiard with four ports and three tunable metasurfaces, shown in Fig.~\ref{FIG_Sup_Sys}(d). The metasurfaces are globally biased with an externally generated DC voltage that can be controlled to the $mV$ level.  The metasurfaces are comprised of 18 unit cell varactor diode-loaded elements arranged linearly, with each cell being $1~\mathrm{cm}$ long. As the bias voltage applied to a metasurface is increased, the effective capacitance of the elements decreases, changing both the amplitude and phase of the reflection spectrum. Because the billiard has a perimeter of $1.574$ meters, each metasurface occupies about $11\%$ of the wall space, and within their tunability range of $8-12~\mathrm{GHz}$ provide a great deal of control over the billiard scattering matrix.

\section{Cavity Shaping and Diabolic Points} \label{Sup_DP}

\begin{figure*} [hbt]
	\centering
	\includegraphics[width=.95\textwidth]{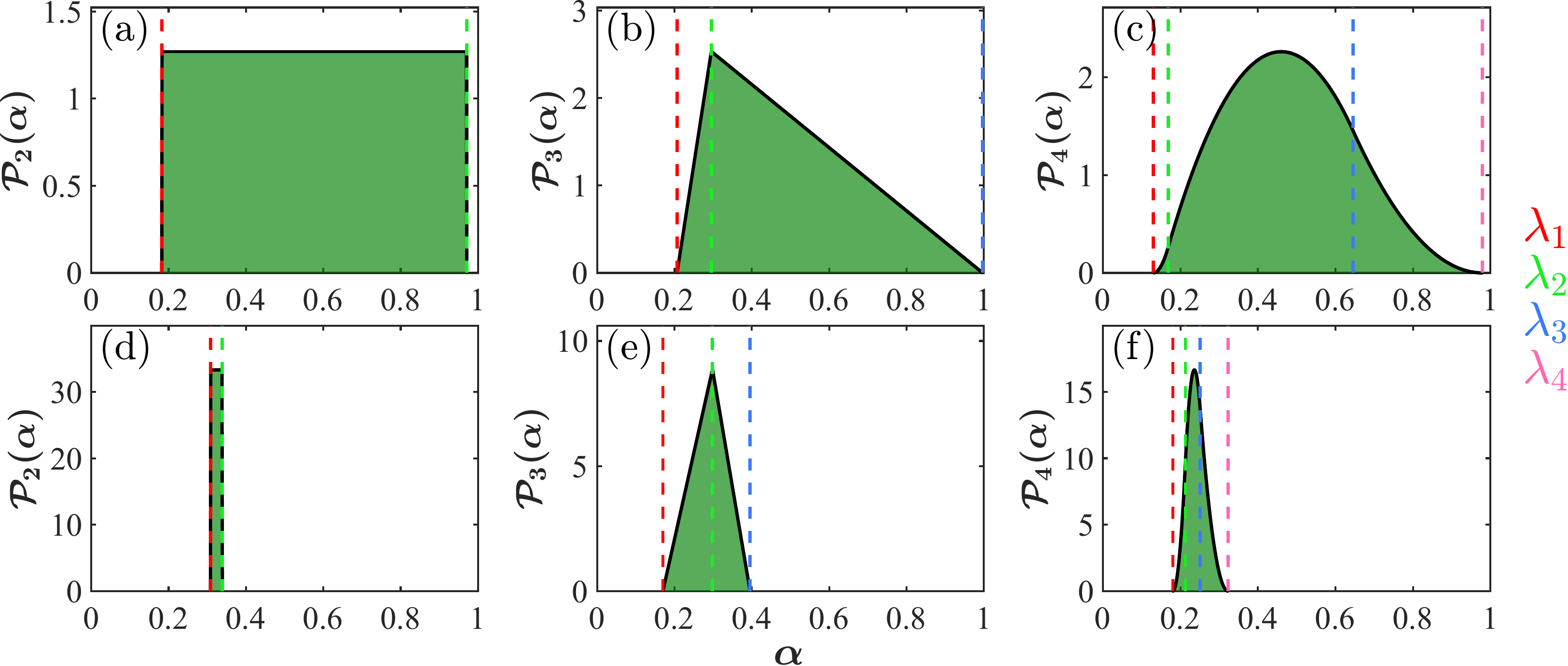}
	\caption{Showing an example of the new cavity shaping paradigm by controlling the variance of absorptivity $\bm{\alpha}$. PDFs are formed from absorptivity matrix $A$ eigenvalues taken from experimental data of microwave Petersen graphs with total length $3.92~\mathrm{m}$. Top (bottom) row shows the PDFs from the phase shifter configurations with the largest (smallest) variance of $\bm{\alpha}$. Panels (a) and (d) come from the $L=2$ channel Petersen graph measured at $7.106~\mathrm{GHz}$, and have $\mathrm{var}[\mathcal{P}_n(\bm{\alpha})]=0.10$ and $\mathrm{var}[\mathcal{P}_n(\bm{\alpha})]=1.5\times10^{-4}$ respectively. Panels (b) and (e) come from the $L=3$ channel Petersen graph measured at $7.791~\mathrm{GHz}$, and have $\mathrm{var}[\mathcal{P}_n(\bm{\alpha})]=0.047$ and $\mathrm{var}[\mathcal{P}_n(\bm{\alpha})]=0.0032$ respectively. Panels (c) and (f) come from the $L=4$ channel Petersen graph measured at $7.136~\mathrm{GHz}$, and have $\mathrm{var}[\mathcal{P}_n(\bm{\alpha})]=0.033$ and $\mathrm{var}[\mathcal{P}_n(\bm{\alpha})]=7.5\times10^{-4}$ respectively.}
	\label{FIG_Sup_EXP_Var}
\end{figure*}

In this section, we show more examples of cavity shaping a system such that an observable of interest has the desired PDF. In Fig.~\ref{FIG_2}(a-c) of the main text, we showed how the average absorptivity $\bm{\alpha}$ can be controlled so that a system with a fixed uniform attenuation rate can be made more or less absorbing. In Fig.~\ref{FIG_Sup_EXP_Var}, we demonstrate in the same systems, how the variance of $\mathcal{P}_n(\bm{\alpha})$ can be controlled to make a system have a very large or small range of possible responses to an input excitation. We calculate the variance of the PDFs as the variance of the eigenvalues divided by $n+1$.

Panels (a) and (d) of Fig.~\ref{FIG_Sup_EXP_Var} show the PDFs $\mathcal{P}_2(\bm{\alpha})$ of the $L=2$ port, $3.92~\mathrm{m}$ Petersen graph measured at $7.106~\mathrm{GHz}$, as in Fig.~\ref{FIG_2}(a). Six of the fifteen bonds of the graph are comprised of programmable analog phase shifters that can be tuned to have an effective electrical length between $25$ and $30~\mathrm{cm}$. By ensuring the net length change of any realization is zero, we can create an ensemble of graphs with a nominally fixed loss parameter. However, sub-centimeter changes of the lengths of a subsets of the bonds of the graph can change the absorptivity matrix $A$ eigenvalues from panel (a)'s $0.18$ and $0.97$ to panel (d)'s $0.31$ and $0.34$. This completely changes the absorption experienced by any waves entering this graph. In panels (b) and (e), we show the same things but for the $L=3$ port Petersen graph at $7.791~\mathrm{GHz}$, and panels (c) and (f) have the $\bm{\alpha}$ PDFs from the $L=4$ port Petersen graph measured at $7.136~\mathrm{GHz}$. 

Note that in the top row of Fig.~\ref{FIG_Sup_EXP_Var}, because the $A$ eigenvalues are bounded above by 1 and below by the minimum absorption possible in the system, which for this graph is about $15\%$, the configurations with the largest variances not only have the largest range of absorption response, but also cover almost the full range of the physically possible absorption. We reiterate that these distributions were found without optimization, instead they were identified by choosing the best realizations of an ensemble with the fixed length constraint. 

\begin{figure} [thb]
	\centering
	\includegraphics[width=\columnwidth]{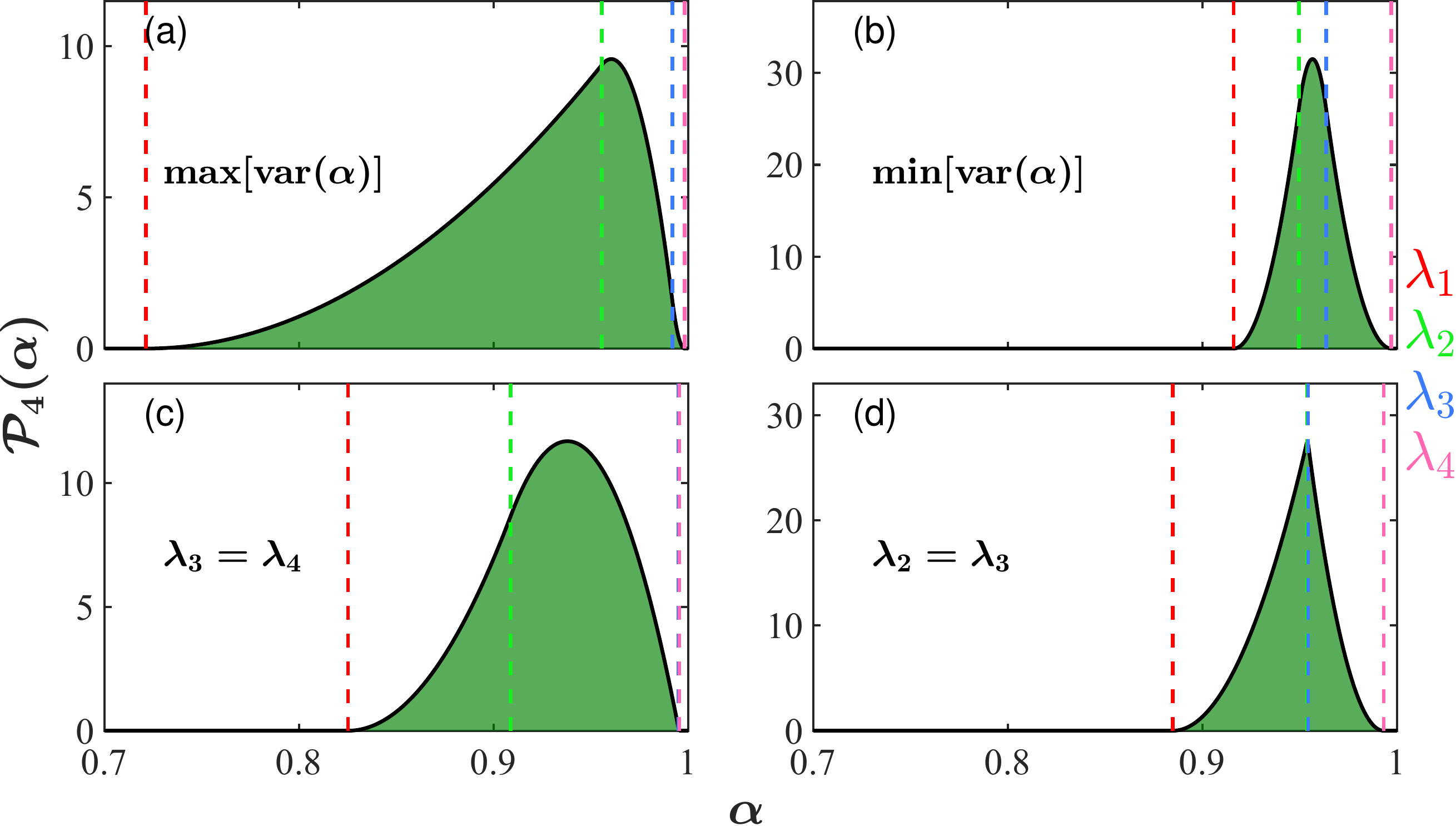}
	\caption{PDFs of absorptivity $\bm{\alpha}$ from the $L=4$ port microwave billiard at $8~\mathrm{GHz}$. For each panel, an optimization function was used to create the condition given in the text: (a) largest variance of $\mathcal{P}_n(\bm{\alpha})$ equal to $0.0035$, (b) smallest variance of $\mathcal{P}_n(\bm{\alpha})$ equal to $2.3\times10^{-4}$, (c) $\lambda_3=\lambda_4$, and (d) $\lambda_2=\lambda_3$. The cases of largest and smallest average $\bm{\alpha}$ for the same system at the same frequency are shown in Fig.~\ref{FIG_2}(d).}
	\label{FIG_Sup_BIL_DP}
\end{figure}

For Fig.~\ref{FIG_Sup_BIL_DP}, we let an optimizer function vary three voltage-controlled metasurfaces within the $L=4$ port quarter bowtie billiard in order to create the desired condition, as we did in Fig.~\ref{FIG_2}(d). All the measurements were performed at $8~\mathrm{GHz}$. In panels (a) and (b), we maximize and minimize the variance of $\bm{\alpha}$, respectively. In panels (c) and (d), we create Diabolic Point (DP) degeneracies of two of absorptivity eigenvalues, which has a profound effect on the shape of the PDF. Because the piecewise polynomials that define the PDF meet at the eigenvalues, by having degenerate eigenvalues we reduce the number of spans (B-spline segments).

Further, at a knot, a B-spline is $p-k$ times differentiable where $p$ is the degree and $k$ is the multiplicity of the knot. If there is DP degeneracy of order $k$, then the PDF is actually only $n-2-k$ times continuously differentiable at the degenerate eigenvalues, as opposed to the $n-3$ times continuously differentiable at a knot corresponding to a unique eigenvalue. That's why there is a cusp in Fig.~\ref{FIG_Sup_BIL_DP}(d), as in this case $n-2-2=0$ and the PDF cannot be differentiated at the knot corresponding to the degenerate eigenvalues $\lambda_2$ and $\lambda_3$.

The cavity shaping demonstrations in this section all focused on absorptivity, but the principal is general and will hold for any observable. Absorptivity was chosen because loss is present in most physical systems of interest and therefore it is almost always of interest and concern to control it.

\section{More Experimental PDFs} \label{Sup_EXP}

In this section we present more experimentally generate PDFs where we perform multi-source injection and directly measure the outgoing waves. This section proves that the theory does not just hold for highly absorptive systems such as the quarter bowtie billiard used in the main text. We also show the experimental PDFs in two different situations, large and small variance, demonstrating that the cavity shaping ideas discussed in the main text and Sec.~\ref{Sup_DP} are sound.

\begin{figure*} [thb]
	\centering
	\includegraphics[width=.95\textwidth]{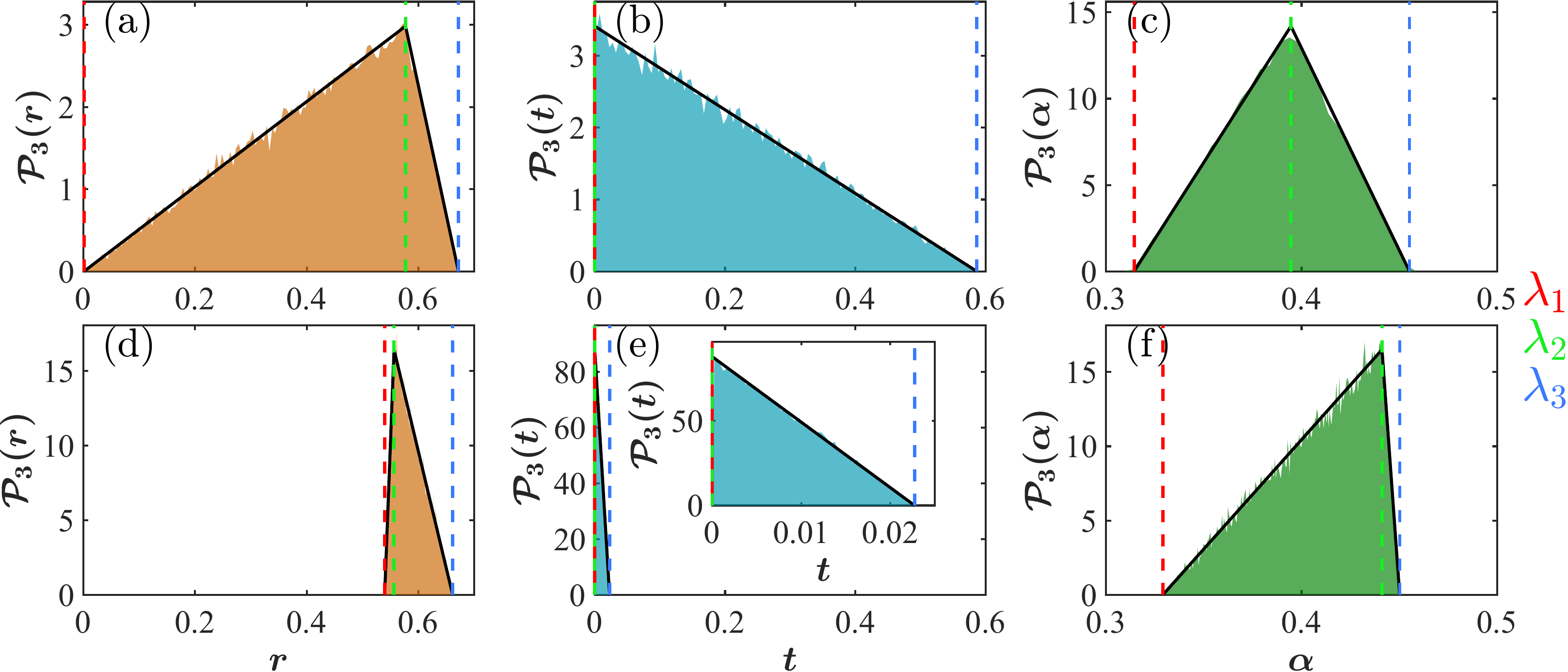}
	\caption{(a-c) Experimental PDFs of reflectivity $\bm{r}$, transmissivity $\bm{t}$ and absorptivity $\bm{\alpha}$ of a $L=4$ port, $n=3$ input channel microwave Petersen graph measured at $1.2715~\mathrm{GHz}$. (d-e) The same PDFs as the row above, but now measured at $1.1283~\mathrm{GHz}$. The locations of the eigenvalues of $R$, $T$, and $A$ are marked in vertical dashed lines in ascending order: red, green, blue.}
	\label{FIG_Sup_EXP_PDF}
\end{figure*}

The experimental platform used in this section is the $L=4$ port, $n=3$ input channel Petersen graph with a total electrical length of $2.38~\mathrm{m}$ measured for the data in Fig.~\ref{FIG_App1}(b), and shown in Fig.~\ref{FIG_Sup_Sys}(b). No phase shifters are included in this graph, so it is not tunable, but by measuring at different frequencies we can see different distributions for the eigenvalues of $R$, $T$, and $A$. In the top row of Fig.~\ref{FIG_Sup_EXP_PDF}, we show $\mathcal{P}_3(\bm{r})$, $\mathcal{P}_3(\bm{t})$, and $\mathcal{P}_3(\bm{\alpha})$ measured at $1.2715~\mathrm{GHz}$. At this frequency, the variance of both $\bm{r}$ and $\bm{t}$ is relatively large, with $0\leq\bm{r}\leq0.672$ and $0\leq\bm{t}\leq0.586$. However, when we measure at $1.1283~\mathrm{GHz}$ (Fig.~\ref{FIG_Sup_EXP_PDF} bottom row), which is just over two mean mode spacings away, we find the variance of $\bm{r}$ and particularly $\bm{t}$ have decreased significantly. Notably, $\mathcal{P}_3(\bm{\alpha})$ doesn't change as much between panels (c) and (f). The center eigenvalue changes from $\lambda_2=0.395$ to $\lambda_2=0.441$, but the extremal eigenvalues hardly move: $\lambda_1=0.314\rightarrow\lambda_1=0.329$ and $\lambda_3=0.455\rightarrow\lambda_3=0.450$.

This tells us that purely spectral shaping (changing frequencies) is less effective for controlling the PDF of $\bm{\alpha}$ than cavity shaping. This makes sense as cavity shaping should have a much more pronounced effect on the interference conditions of wave excitations within the system. Destructive interference at the channels and constructive interference at sources of loss are instrumental for increasing the absorptivity experience by a wavefront without directly tuning the lossiness of the system. While spectral shaping of course also effects wave interference by changing the wavelength, it is in a much smoother and global manner which offers less control. Therefore with spectral shaping one is much more constrained by the intrinsic attenuation rate of the system.

\clearpage
\newpage

\bibliography{Refs.bib}






\end{document}